\documentclass{elsart}
\usepackage{amsfonts}

\usepackage{amsmath}

\input{tcilatex}

\begin{document}

\begin{frontmatter}

\title{Entropic Nonextensivity as a Measure of Time Series Complexity}

\author{Milan Rajkovi\'c}
\address{Institute of Nuclear Sciences Vin\v ca,  
   P.O. Box 522, 11001 Belgrade, Serbia }

\begin{abstract}
Information entropy is applied to the analysis of  time series generated by dynamical systems. Complexity of a 
temporal or spatio-temporal signal is defined as the difference between the sum of entropies of the local linear 
regions of the trajectory manifold, and the entropy  of the globally linearized manifold. When the entropies are 
Tsallis entropies the complexity is characterized by the value of $q$.   
\end{abstract}


\end{frontmatter}

\section*{Introduction}

Detection of phase changes in a dynamical system and classification of
complex signals based on their specific features represent important and
challenging issues in signal processing applied to as diverse areas as fluid
dynamics, climate prediction, atmospheric science, astronomy,
electrophysiology etc. In order to characterize signals by their intrinsic
and invariable characteristics we pursue approach based on the study of
topological properties of the manifold obtained from either a single or
multiple (i.e. spatio-temporal) signals, previously described in \cite{MR},
and references therein. The starting point of the method is reconstruction
of the trajectory manifold, based on the Takens embedding theorem \cite%
{Takens},\cite{Packard} which, it should be emphasized, is concerned with
purely deterministic dynamical systems. However it is possible to extend
Taken's framework both to deterministically forced systems and to stochastic
systems\cite{Stark}. We define complexity of temporal or spatio-temporal
signal as the difference between the sum of entropies of locally linear
regions of the manifold and the entropy of the linearized whole topological
entity (the trajectory manifold or the attractor). To each local and global
linear manifold corresponds a matrix whose eigenvalues define probabilities
figuring in the entropy expressions for which we assume normalized Tsallis
entropy form. The value of nonextensivity parameter $q$, figuring in the
expression for Tsallis entropy, is determined so that it corresponds to the
(first) maximum value or to the (first) minumum value of complexity. The
former corresponds to the case when $q>1$ and the latter to the case $q<1$.
Hence $q$ assigns a unique complexity quantifier to each temporal or
spatio-temporal signal (i.e. to a particular dynamical state represented by
a given signal). Moreover, complexity defined in such a manner gives a
specific physical meaning to the nonextensivity parameter $q$ as a degree of
signal's essentially topological complexity. The organization of the paper
is as follows: In section 1 we present the essential features of local
linear analysis of the trajectory manifold and the method of global
linearization of the manifold. We also define the corresponding local and
global Tsallis entropies. In section 3 we present an application of the
method for detecting differences in a temporal evolution of the signal of
the electrical activity of the human brain (EEG) preceeding the onset of the
epileptic seizure. Finally, we discuss the relevance and possible variations
of this signal analysis framework for future applications.

\section{Entropy of local and global linear trajectory manifold}

In case of spatio-temporal signals the embedding space is constructed with
reference to the spatial dependence\cite{Torcini} while a delay-time method
is used in case of temporal signals\cite{Packard}. In the latter case delay
time is determined using the first minimum of the auto mutual information
function\cite{Frazier}, while the embedding dimension is usually determined
based on certain topological criteria, such as the method of Cao\cite{Cao}.
The goal of the time-series analysis is to learn as much as possible about
the underlying system given only the time series of a single experimentally
or numerically obtianed quantity that is a function of the state of the
system. The embedding procedure yields a manifold $\mathcal{A}$ which
locally looks like $\mathbb{R}^{n}$, where $n$ is the dimension of $\mathcal{%
A}$. The dynamics in phase space is followed in the set of local coverings
of the manifold $\mathcal{A}$ obtained by the embedding (reconstruction)
procedure, such that the dynamics is projected locally into the tangent
space at various points of the manifold. In case of chaotic dynamical
systems a trajectory may be followed on the attractor instead of the
manifold, although our method does not require asymptotic dynamics (i.e. a
large number of data points) to extract information from the time (or
space-time) series. A given point on the manifold, defined by the vector $%
\overset{\rightarrow }{x}_{1},$ is chosen as the center of a hypersphere of
radius $\epsilon $, consisting of the points $\overset{\rightarrow }{y_{i}}=%
\overset{\rightarrow }{x_{i}}-\overset{\rightarrow }{x}_{1}$, such that $||%
\overset{\rightarrow }{y_{i}}||$ $\leq \epsilon $. The radius of the
hypersphere (i.e the number of points) is determined so that the local
region of the manifold is linear\cite{MR}. The points contained in this
hypersphere are represented by the matrix $B$, whose rows are approximately
tangent vectors to $\mathcal{A}$ at $\overset{\rightarrow }{x}_{1}.$ On this
matrix a singular value decomposition (SVD) is performed and the orthogonal
complement, the Gaussian component (noise), is extracted first\cite{MR}. In
case of a spatio-temporal signal the next step may be the separation of
active and passive modes which belong to two closed, mutually orthogonal
linear subspaces, $\mathcal{F}_{a}$ and $\mathcal{F}_{s}$ respectively. The
coordinates of directions in the orthogonal space $\mathcal{F}_{s}$ (slaved
modes) are functions of coordinates in the tangent space $\mathcal{F}_{a}$
(active modes), and the dynamics is completely determined and cotrolled by
the dynamics on $\mathcal{F}_{a}$. Noise separation is performed using
information theoretical criteria or using a method based on the perturbation
of singular values \cite{MR}, depending on the system. However, in case of
neurophysiological signals noise is not\ removed since it may represent an
intrinsic part of the signal, carrying important information about the
underlying dynamics. Following filtering of the Gaussian component, the SVD
of the local dynamics may be represented in matrix form%
\begin{equation}
S=U\Sigma V^{H}=(U_{a}U_{s})\left( 
\begin{array}{ll}
\Sigma _{a} & 0 \\ 
0 & \Sigma _{s}%
\end{array}%
\right) \left( 
\begin{array}{c}
V_{a}^{\ast } \\ 
V_{s}^{\ast }%
\end{array}%
\right) =\sum\nolimits_{i=1}^{r}\sigma _{i}u_{i}v_{i}^{H}
\end{equation}%
where $u_{i}$ and $v_{i}$ are the orthonornmal characteristic vectors of the
matrix $BB^{T}$ (or $B^{T}B$) and \{$\sigma _{i}$\} are the corresponding
characteristic values. In the above expression the matrices are assumed to
be real. Indices $a$ and $s$ refer to active and slaved modes (subspaces),
hence the above expression originates from the analysis of spatio-temporal
dynamics. Index $r$ represents the rank of the matrix $B$ and since noise
has been extracted it is actually the number of points, $N_{L}$, contained
in the local region of the manifold. It is important to emphasize that since
matrix $B$ contains information about the local curvature of the manifold it
contains essential information about the (local) dynamics of the system
under consideration. Dynamics (flow) is followed on the manifold $\mathcal{A}
$ so that, depending on the choice of the reference point, either
overlapping or nonoverlapping local linear regions may be obtained, however
in further exposition we assume nonoverlapping local regions. The
eigenvalues \{$\lambda _{i}$\} (squares of singular values \{$\sigma _{i}$%
\}) are used to construct an information entropy for each local linear
region which we introduce as the Tsallis normalized entropy\cite{Tsallis}%
\begin{equation}
(S_{q}^{L})_{j}=\frac{\sum\nolimits_{i=1}^{N_{L}}p_{i}^{q}-1}{N_{L}^{1-q}-1},
\end{equation}%
where index $j$ corresponds to the $j$-th local region, and where
probabilites $p_{i}$ are defined as%
\begin{equation}
p_{i}=\frac{\lambda _{i}}{\sum\nolimits_{i=1}^{N_{L}}\lambda _{i}}.
\end{equation}%
In global linearization, the usually high-dimensional manifold corresponding
to the temporal or spatio-temporal signal is deformed until it is absorbed
by a linear subspace. Briefly, the algorithm\cite{MR} consists in forming a
minimum spanning tree (MST) as the structure invariant of the original
configuration of points, whose length remains constant during the
linearization procedure. Next the actual linearization mapping, the
barycentric transformation, is applied iteratively and after each iteration
the MST length is restored. The procedure is continued until the complete
structure is linearized. Singular values are computed for the final
configuration and the number of dominant singular values is determined,
after the removal of the noise subspace. We define the normalized global
entropy of the signal as%
\begin{equation*}
S_{q}^{G}=\frac{\sum\nolimits_{i=1}^{N_{G}}p_{i}^{q}-1}{N_{G}^{1-q}-1},
\end{equation*}%
where $N_{G}$ is the total number of points in the signal after noise
removal, and where probabilities are defined by the eigenvalues \{$\mu _{i}$%
\} of the matrix corresponding to the global linear manifold%
\begin{equation*}
p_{i}=\frac{\mu _{i}}{\sum\nolimits_{i=1}^{N_{G}}\mu _{i}}.
\end{equation*}

\section{Complexity of the signal}

We define the complexity measure as the difference between the sum of local
entropies and the entropy.of the globally linearized manifold: 
\begin{equation*}
C=\sum\nolimits_{j}(S_{q}^{L})_{j}-S_{q}^{G}.
\end{equation*}%
Clearly $C$ as the sum and difference of entropies has all the properties
that characterize the Tsallis entropy. Since dominating modes determining
entropy expressions also deterimine local (or global) topological dimension
of the manifold $\mathcal{A}$, complexity measure $C$ reflects the balance
of local and global topological properties of the trajectory manifold and
the way the energy of the signal (reflected in the eigenvalues) is
distributed (locally and globally) on the manifold. Moreover, $C$ may
represent a measure of organization or self-organization of the dynamical
system, depending whether the external agent imposing the organization
exists or not. As the system organizes or self-organizes, the entropy of the
linearized whole manifold $\mathcal{A}$ decreases based on the information
exchanged between local regions of the trajectory manifold. Assuming $q>1$,
since entropies are normalized the maximum possible value of $C$ is equal to
the number of local linear regions on the manifold, say $M$. As $q$
increases, the value of $C$ will tend to $M$, and the (first) value of $q$
corresponding to the case when $C$ becomes maximal represents unique
complexity measure of the temporal or spatio-temporal signal, which we
denote as $q^{\ast }$. Hence, 
\begin{equation}
q^{\ast }=(q\mid C=\max (\sum\nolimits_{j}(S_{q}^{L})_{j}-S_{q}^{G})).
\end{equation}%
The value of $q\ast $ is very sensitive to variations of either one of the
probability values $p_{i}$ and $\mu _{i},$ and hence on the dynamical and
topological features of the dynamical system. \ It also represents a more
precise and delicate characterization of the signal (dynamical system) then
just the value of $C$ alone. \bigskip

\section{An example}

In order to illustrate the use of complexity measure introduced here, we
choose an electroencephalogram (EEG) time signal recorded few minutes before
the epileptic seizure. Normally, rich thalamocortical (TC)-CT feedback loops
regulate the flow of information to the cortex and this ability is
transiently lost in absence seizures, because large numbers of CT loops are
engaged for seconds in much stronger, low-frequency ( approximately 3 Hz)
oscillations\cite{Kost}. Signal $a$ in Fig. 1 represents the portion of the
signal recorded 2 min. before the seizure, while $b$ shows the onset of
seizure which lasted approximately 10 sec. Assuming $q>1$, we compare the $%
q^{\ast }$ values of the signal comprising 6000 points (sampled at frequency
of 128Hz) recorded 2 min before the seizure (signal 1), the signal of the
same duration recorded 45 sec. before the seizure (signal 2), and the
portion of the signal 30 sec. before the event (signal 3). For signal 1 the
obtained value is $q^{\ast }=3.6$ (graph $a$ in Fig. 2) while for signal 2, $%
q^{\ast }=2.7$ (graph $b$ in Fig. 2). For signal 3, $q^{\ast }$ is equal to $%
2.1$. Hence, there is a clear indication of the emergence of pattern in the
time-series before the actual occurence of the seizure. It should be
emphasized that conventional methods of nonlinear dynamics such as
correlation dimension, Lyapunov exponents or Kolmogorov enropy are not able
to provide such conlcusive evidence for the change in the dynamical features
of the EEG signal. We speculate, based on our previous work\cite{MR}, that
by monitoring the evolution of complexity of a certain dynamical systems $%
q^{\ast }$ value may be of valuable practical importance.

\section{Conclusion}

A new method for detection of changes in the evolution of dynamical systems
as well as for classification of temporal and spatio-temporal signals is
presented. The method relies on the nonextensive Tsallis entropy as the
basic tool in forming the complexity measure which may prove to be of
practical importance for both analysis as well as predictive pruposes.

\section{Acknowledgements}

This work was partially supported by SMST project OI 1986.

\bigskip

\bigskip FIGURE\ CAPTIONS

Fig. 1 a) EEG signal recorded at spatial position C4 2 min.before the onset
of epileptic seizure; b) the onset of the seizure is characterized by large
amplitude bursts of low frequency (\symbol{126}3Hz) Sampling frequency is
128 Hz. Signals were filtered with anti-aliasing low-pass filter.

\bigskip

Fig. 2 a) Complexity as a function of parameter q for the signal recorded 2
min before the seizure onset; b) same for the signal recorded 45 sec. before
the onset of the seizure.

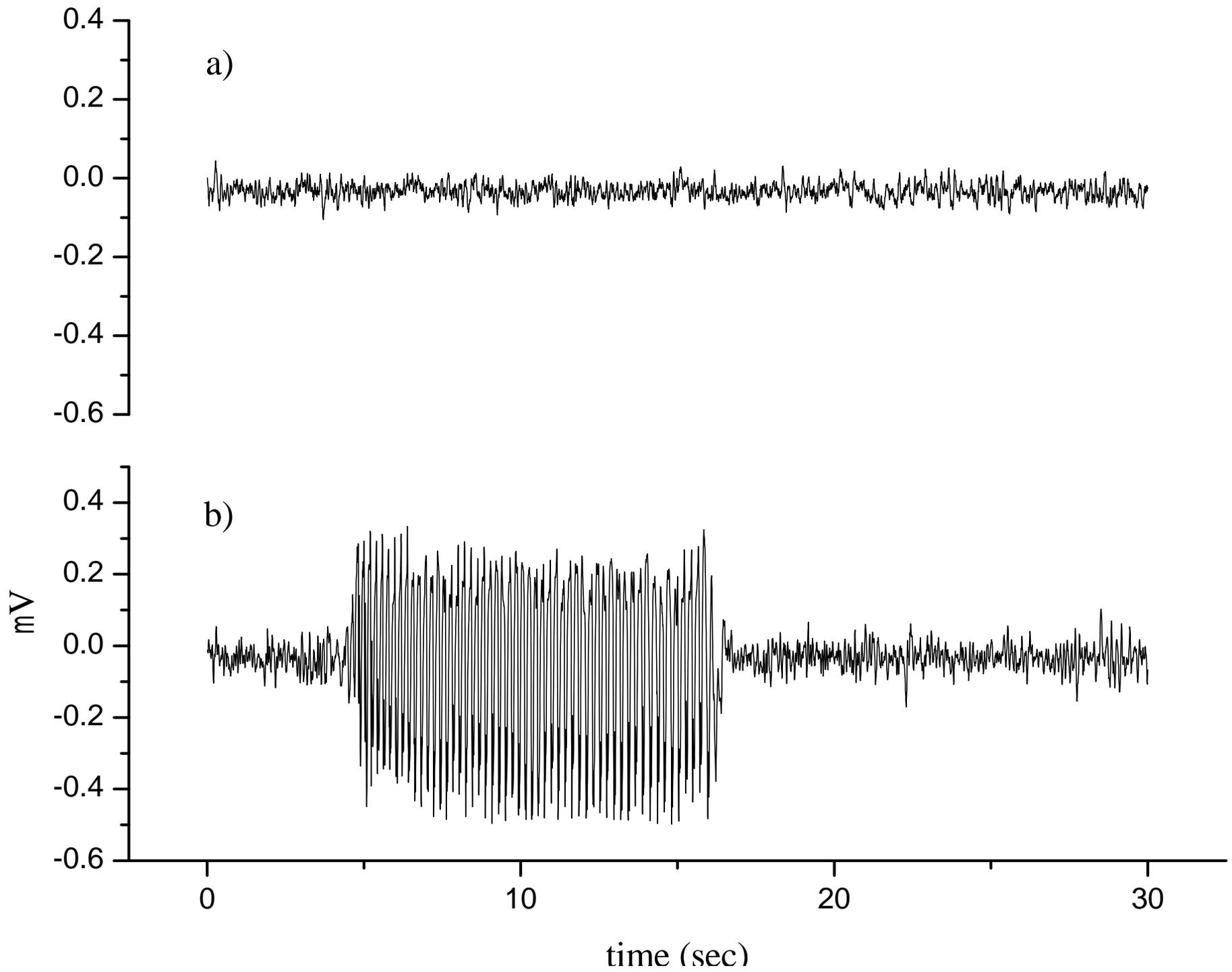
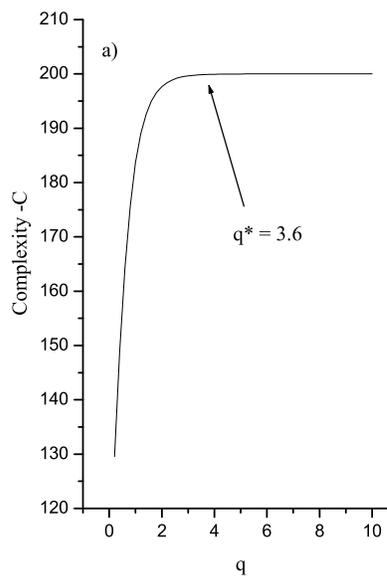
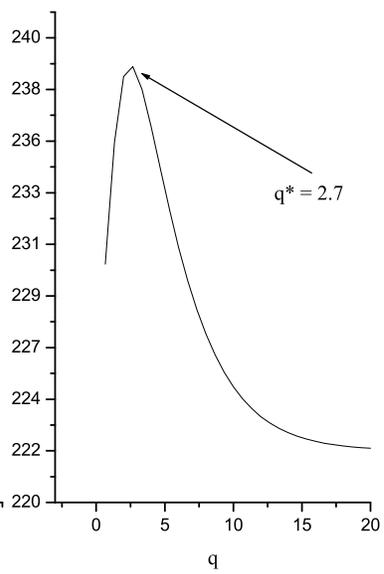
\end{document}